\documentclass[aps,twocolumn,showpacs,preprintnumbers,nofootinbib]{revtex4-1}

\usepackage{graphicx}  

\usepackage{fancyhdr}
\usepackage{parskip}
\usepackage{dcolumn}   

\usepackage{bm}        
\usepackage{amsfonts}  
\usepackage{amsmath}   
\usepackage{amssymb}   

\newcommand{\pwisein}{\left\{ \begin{array}{ll}}
\newcommand{\pwiseout}{\end{array}\right.}

\begin{document}

\title{Yb fiber amplifier at 972.5 nm with frequency quadrupling to 243.1 nm}

\author{Z. Burkley$^{1,\ast}$}\footnote{Corresponding author: zakary.burkley@colostate.edu}
\author{C. Rasor$^1$}
\author{S. F. Cooper$^1$}
\author{A. D. Brandt$^1$}
\author{D. C. Yost$^1$}
\vspace{10pt}
\affiliation {$^1$Department of Physics, Colorado State University, Fort Collins, CO, 80521}

\begin{abstract}  

We demonstrate a continuous-wave ytterbium-doped fiber amplifier which produces 6.3 W at a wavelength of 972.5 nm.  We frequency quadruple this source in two resonant doubling stages to produce 530 mW at 243.1 nm.  Radiation at this wavelength is required to excite the 1S-2S transition in atomic hydrogen and could therefore find application in experimental studies of hydrogen and anti-hydrogen.
\end{abstract}

\maketitle 

\section{Introduction}
\label{intro}
The hydrogen 1S-2S two-photon transition was first observed by H\"{a}nsch {\it et. al} in 1975 \cite{Hansch:75}. Over the following four decades, the continued improvement in the spectroscopy of this transition has offered stringent tests of quantum electrodynamics (QED) and has led to increasingly precise determinations of the Rydberg constant and proton charge radius \cite{Mohr:12}.  The importance of the 1S-2S transition stems in part from the simplicity of hydrogen, which makes it amenable to theoretical study, and also from the its narrow natural linewidth, which is only 1.3 Hz.

When reviewing the well-known measurements of the 1S-2S transition, one can also observe a continual refinement of the spectroscopy lasers used -- first by a transition from pulsed to cw lasers \cite{Couillaud:84, Foot:1985} and then by a continued increase in power, coherence and robustness \cite{Boshier:89, Kallenbach:91, Zimmerman:95, Kolachevsky:06, Kolachevsky:11}.  The most recent result was reported by the H\"{a}nsch group in 2011, in which they determined the transition to a fractional frequency uncertainty of $4.2 \times 10^{-15}$ \cite{Parthey:11}.  By that time, the UV laser source had evolved to an all solid state system that produced 13 mW of 243 nm cw radiation.  This radiation was cavity enhanced to 368 mW within the hydrogen spectrometer.  In addition to the impressive intracavity power, this radiation source possessed an extremely narrow linewidth of $\approx$ 1 Hz which is commensurate with the hydrogen 1S-2S transition width itself.  More recently, in 2013, Beyer et al. reported on a 243 nm laser which was capable of producing 75 mW before cavity enhancement \cite{Beyer:13}.

Notwithstanding these accomplishments, we believe that continuing to increase the laser power at 243 nm would be very beneficial.  For instance, the 1S-2S transition could be excited with laser beams of large transverse dimensions which could decrease transit-time broadening and increase the proportion of atoms in the atomic beam which are excited. With the recent trapping of anti-hydrogen in its ground state, a larger beam would also prove beneficial in mitigating the difficulties created by the low number of trapped anti-hydrogen atoms available \cite{Gabrielse:12, Andersen:11}. However, we are mainly motivated to develop a power scalable 243 nm laser in order to explore proposals to laser cool atomic hydrogen using the 1S-2S transition \cite{Zehnle:2001, Kielpinski:06, Wu:11}. 

Spectroscopy of hydrogen and the recently trapped anti-hydrogen would benefit tremendously from robust laser cooling. Two photon laser cooling could be more rapid and flexible than the more traditional approach using Lyman-alpha radiation at 121.6 nm -- mostly due to the greater ease at producing radiation at 243 nm.  To obtain reasonable scattering rates with such schemes requires that the 2S state is coupled to a state with short lifetime, for instance either the 2P \cite{Kielpinski:06} or 3P \cite{Wu:11} states, and the average power of the cavity enhanced 243 nm radiation source should be at the $\sim$100 W level.  For a beam diameter of $\sim$ 500 $\mu$m, this would lead to a scattering rate of $\sim$500 Hz when maximally coupling the 2S and 2P states \cite{Kielpinski:06}.  Power enhancement within an optical cavity can reach factors of $\sim$100 with commercially available mirrors so that Watt-level 243 nm sources could be sufficient for an initial demonstration of cooling.

Here, we present a laser system which is a major step towards laser cooling hydrogen with the two-photon 1S-2S transition. The system is composed of an extended cavity laser diode (ECDL) at 972 nm followed by a tapered amplifier, a Ytterbium-doped double clad fiber amplifier, and two consecutive resonant doubling stages.  The Ytterbium (Yb) fiber amplifier is a notable feature of this source since gain is much more readily obtained in Yb systems at $\sim$1030 nm due to the low absorption cross section at this wavelength. Gain near the emission cross section peak at 976 nm is also possible but requires population inversions above $\sim$50\% because the absorption cross section in that spectral region has approximately the same magnitude. Despite this difficulty, there have been demonstrations of 100 W Yb doped fiber lasers near the emission cross section peak at 976 nm \cite{Boullet:08, Roser:08}. To the best of our knowledge, there have been only a few Yb fiber-based laser systems which operate at shorter wavelengths and these produced relatively low power ($\sim10$ mW) \cite{Hanna:90, Yi:12}. The Yb fiber amplifier we demonstrate here produces 6.3 W of power at 972 nm which upon frequency quadrupling gives 530 mW of power at 243 nm. We believe this approach is power-scalable and that we can continue to increase our UV radiation power by scaling the fundamental power at 972 nm. 

\section{Seed Laser and Yb Fiber Amplifier}
\label{amplifier}
\begin{figure}[h]
\centering
\includegraphics[width=\linewidth]{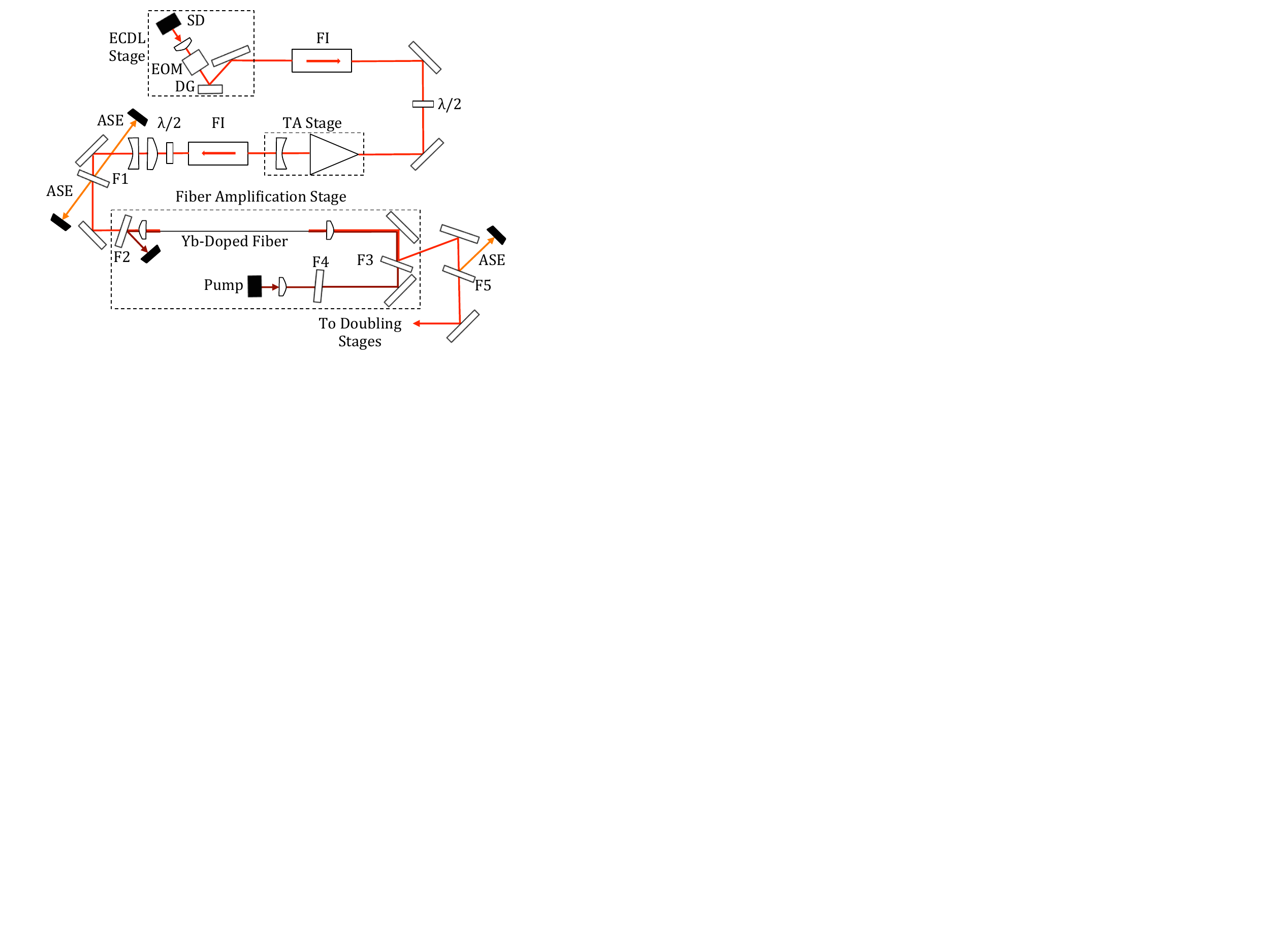}
\caption{Schematic of the ECDL Master Oscillator and Amplification Stages. SD: Seed Diode, DG: Diffraction Grating, FI: Faraday Isolator, F1 and F5: Bandpass Filters, F2: Longpass Filter, F3 and F4: Shortpass Filters. The ECDL contains an electro-optic modulator (EOM) for fast frequency control although it was not used for the studies here.}
\label{fig:setup}
\end{figure}

Our master oscillator is an external cavity diode laser (ECDL) that follows the design in \cite{Kolachevsky:11}, in which the length of the cavity was increased to 20 cm. This reduced the laser linewidth by a factor of 10 in comparison to an ECDL with a more typical 2 cm long cavity. We chose a somewhat shorter cavity length of 10 cm to increase the mode-hop free tuning range.  The ECDL produces 30 mW of power at 972 nm and the design is shown in Fig. \ref{fig:setup}.  The output of the oscillator is amplified to 3 W with a commercial tapered amplifier (TA).  The TA produces a complex mode structure and only $\approx$ 2.4 W is contained within a TEM$_{00}$ mode.  

The output from the TA is then further amplified within a double-clad Yb doped fiber with a 20 $\mu$m diameter core and 128 $\mu$m cladding. The core of this fiber has a numerical aperture of 0.075 which is large enough to support the propagation of a few higher-order modes. With careful alignment, we can achieve a TEM$_{00}$ beam at the output of the amplifier.

To obtain the necessary population inversion requires that a high pump intensity at 915 nm be maintained along the entire length of the fiber, which results in low pump absorption.  In our case, we use a short section of fiber ($\sim$ 10 cm) which absorbs only $13 \%$ of the incident pump power.   Due to the high population inversion within the fiber amplifier, there is also significant gain within the 1030 nm and 976 nm spectral regions. This is problematic as significant amplified spontaneous emission (ASE) would degrade the amplifier performance by reducing the population inversion.  As shown in \cite{Nilsson:98}, the gain at a given wavelength within a homogeneously broadened amplifier can be written as a function of the gain or absorption at two other wavelengths and their respective absorption and emission cross sections. Using the cross section data for our fiber, we find an expression for the gain ($G_{\lambda}$) at 976 nm given by
\begin{equation}
G_{976}= 2.28 \cdot G_{972} + 1.04 \cdot  \beta A_{915}.
\end{equation}
Here $A_{915}$ is the absorption of the pump at 915 nm, and $\beta$ is the ratio of the cladding area to the core area. In our amplifier, $\beta = 40$, $G_{972} =4.2$ dB and  $A_{915}=0.6$ dB which results in a gain of $G_{976} = 35$ dB.  Similarly, the gain at 1030 nm where Yb doped fiber lasers commonly operate is given by
\begin{equation}
G_{1030}= 0.49 \cdot G_{972} + 1.28 \cdot \beta A_{915},
\end{equation}
which results in $G_{1030} = 32$ dB.  As can be seen from the previous expressions, the gain at 976 nm and 1030 nm depends sensitively on the pump absorption due to the large value of $\beta$.  As discussed extensively in \cite{Nilsson:98, Boullet:08, Roser:08}, increasing the pump absorption, and therefore efficiency of the amplifier, would need to be accompanied by a decrease in $\beta$ to keep the gain at around 976 nm and 1030 nm manageable.  Even in our current configuration, the $= 32$ dB gain at 1030 nm would cause the amplifier to lase if the ends of the gain fiber were flat cleaved.  To mitigate these effects, we angle polish the ends of the gain fiber and use bandpass filters (FWHM$=$4 nm) to remove ASE originating from both the fiber amplifier and the TA \cite{Yi:12}.

\begin{figure}[h]
\centering
\includegraphics[width=\linewidth]{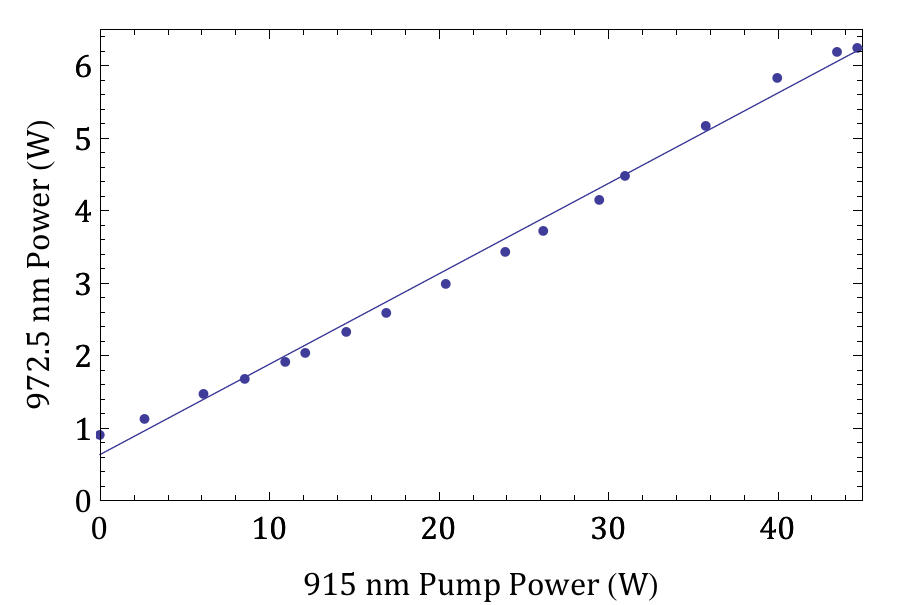}
\caption{Fiber Amplified 972 nm radiation as a function of 915 nm pump power. The data shown is with 3 W of seed power incident on the fiber. With 45 W of pump power, approximately 6.3 W of 972 nm radiation is produced from the fiber amplifier.  The slight deviation from the linear trend line is due to a shift of the pump to longer wavelengths as the power is increased.}
\label{fig:amp}
\end{figure}

The high inversion in the fiber can also lead to photodarkening -- a poorly understood decrease in the optical transmission of gain fibers which degrades performance \cite{Koponen:06}. Mitigation of this effect is possible by codoping the gain fiber with phosphorous or cerium \cite{Engholm:08, Engholm:09}.  We use a fiber codoped with phosphorous because it was commercially available and we have yet to observe any such degradation of the amplifier performance due to this effect.

The output power of the fiber amplifier as a function of 915 nm pump power is shown in Fig. \ref{fig:amp}. At our maximum pump power of 45 W, we obtain an output power of 6.3 W at 972 nm.  As seen in Fig. \ref{fig:amp}, there is a near linear increase in the output power as a function of pump power.  The small deviation from the linear trend line is due to an increase of the wavelength of the pump radiation as the diode current is increased.  We have modeled the 972 nm power, 915 nm pump power and ASE along the length of the fiber based on the method found in \cite{Roser:08}.  The model indicates that we can continue to linearly increase the output power with additional pump power. These same models suggest an additional  amplification stage will also be an effective means to increase the output power \cite{Clarkson:10}.

\section{Doubling Stages}
\label{doubling}
\begin{figure}[h]
\centering
\includegraphics[width=\linewidth]{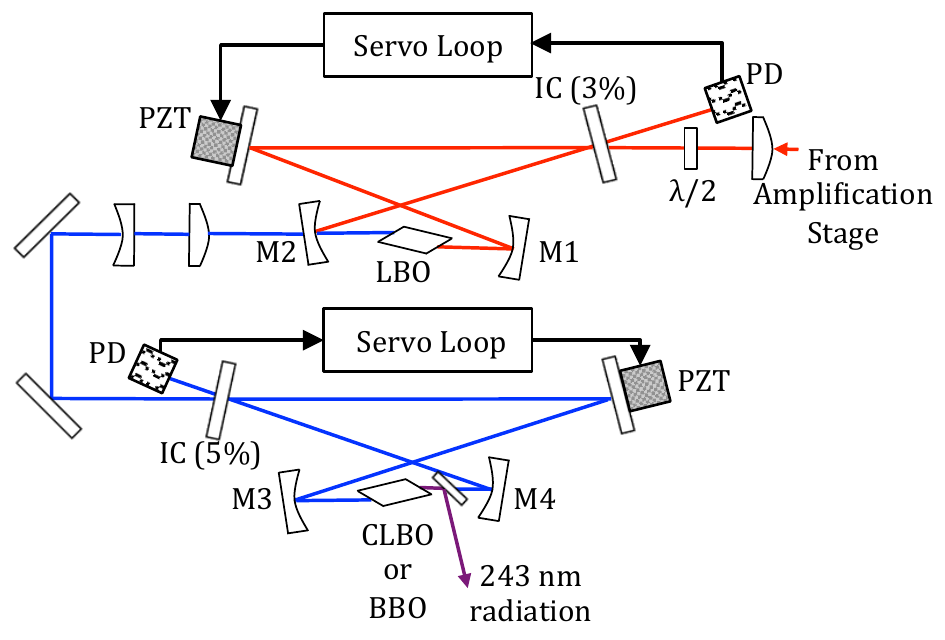}
\caption{Experimental Setup of Doubling Stages. PD: Photodiode, PZT: PiezoElectric Transducer, IC: Input Coupler, M1-M4: 200 mm ROC mirrors.  For the servo loops, we use dither locks to maintain the resonance condition.}
\label{fig:dub}
\end{figure}

As shown in Fig. \ref{fig:dub}, the output of the fiber amplifier is frequency quadrupled to 243 nm in two consecutive resonant doubling stages.  The first stage uses lithium triborate (LBO) as the nonlinear crystal, while for the second we tested both barium borate (BBO) and cesium lithium triborate (CLBO) crystals.  

The first nonlinear doubling stage uses a standard bowtie geometry. The curved mirrors have a radius of curvature (ROC) of 200 mm, producing a beam waist within the LBO of $62 ~\mu$m. This increased mode size, which is $\sim 1.8$  times the optimal waist determined from the Boyd-Kleinman focusing criteria \cite{Boyd:68}, increases the robustness of the doubling stage with minimal effect on the overall conversion efficiency. We use type I non-critical phase matching in order to eliminate spatial walk-off and improve the 486 nm output beam quality, which requires the LBO be kept at a temperature of 283$^\circ$ C.   Because the performance of typical dual wavelength antireflection coatings is not guaranteed at high temperatures, we use a Brewster cut crystal to reduce the loss of the resonant 972 nm light.
 
\begin{figure}[h]
\centering
\includegraphics[width=\linewidth]{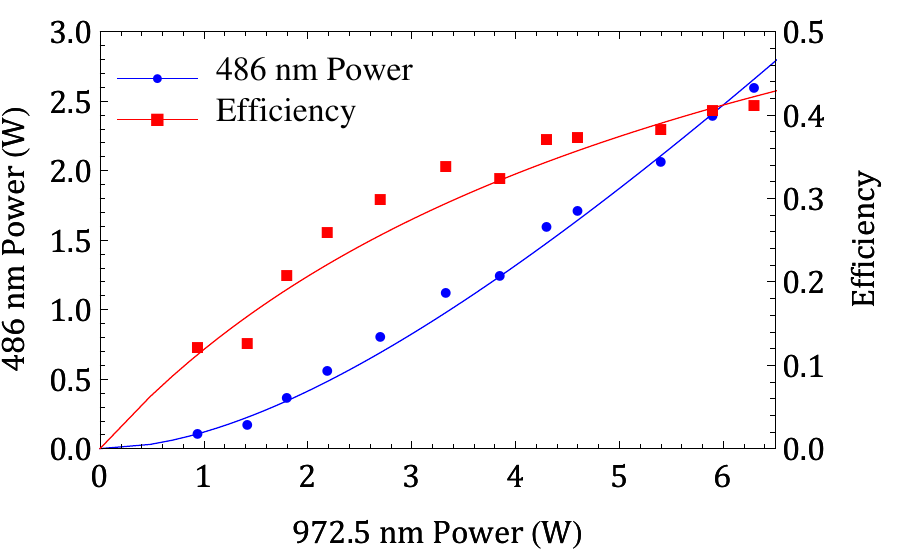}
\caption{Experimental results of frequency doubling fiber amplified 972 nm radiation using LBO as the nonlinear medium. With 6.3 W of power at 972 nm, up to 2.60 W of power at 486 nm is generated. The theoretical fit for harmonic conversion used here follows the model presented in \cite{Polzik:91}.}
\label{fig:LBO}
\end{figure}

This led to an 18\% loss of the generated 486 nm radiation from the Fresnel reflection on the crystal output facet. The remaining 82\% of the 486 radiation was coupled out of the cavity through a dichroic curved mirror with high reflectivity at 972 nm and high transmission ($>90\%$) at 486 nm.  The 486 nm output power as a function of incident fundamental power is shown in Fig: \ref{fig:LBO}.  We obtain 2.4 W of 486 nm radiation with 6.3 W of 972 nm fundamental power.  

\begin{figure}[h]
\centering
\includegraphics[width=\linewidth]{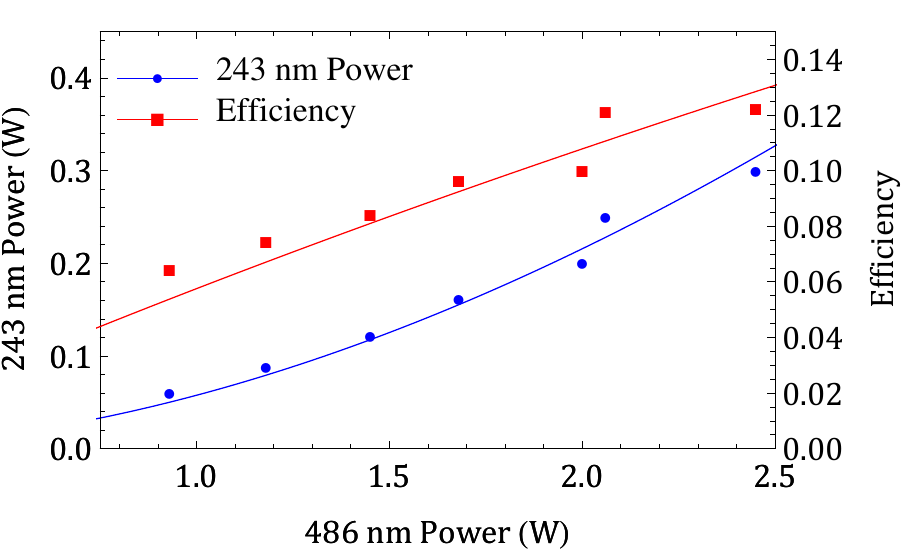}
\caption{Experimental results of frequency doubling 486 nm radiation to 243 nm radiation using BBO as the nonlinear medium. With 2.4 W of 486 nm radiation, up to 300 mW of power at 243 nm is generated.}
\label{fig:BBO}
\end{figure}

\begin{figure}[h]
\centering
\includegraphics[width=\linewidth]{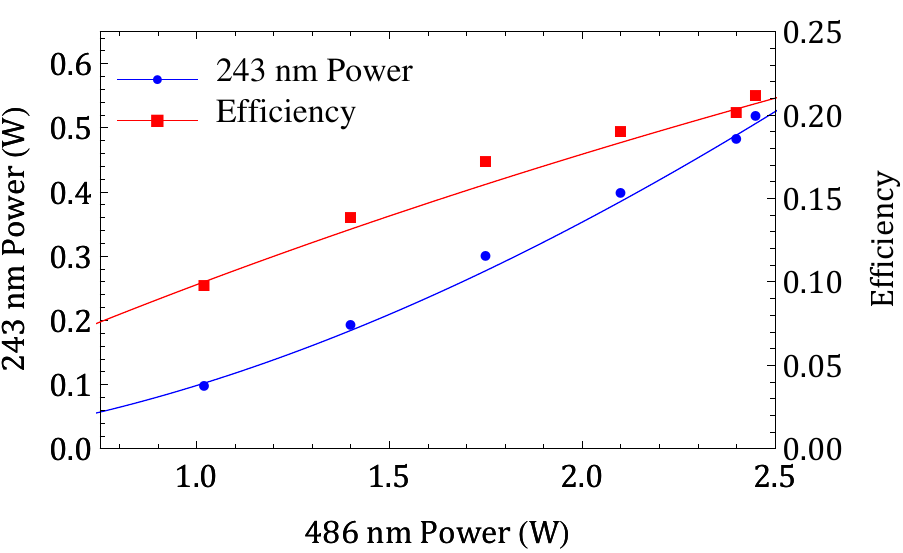}
\caption{Experimental results of frequency doubling 486 nm radiation to 243 nm radiation using CLBO as the nonlinear medium. With 2.45 W of 486 nm radiation, up to 530 mW of power at 243 nm is generated.}
\label{fig:CLBO}
\end{figure}

In the second doubling stage, which produces 243 nm radiation, we tested both a BBO crystal and a CLBO crystal \cite{Scheid:07, Sakuma:04, Hu:13, Kaneda:08} in type I critical phase matching configurations. CLBO has a lower nonlinear coefficient than BBO, but also less spatial walk-off and a higher damage threshold \cite{Mori:95}. Similar to the first stage, the second doubling stage is a bow tie design with 200 mm ROC mirrors to produce a focus in the nonlinear crystal. This produces a beam waist of 44 $\mu$m which is $\sim 1.9$ times the Boyd-Kleinman focusing criteria \cite{Boyd:68} in order to prevent damage of the nonlinear crystal at high intensities and to minimize walkoff effects. Both crystals were Brewster cut and 10 mm long.  In this case, Brewster cut crystals were used because AR coatings are not yet well-developed for CLBO crystals \cite{Sakuma:04}.  This introduced a 27\% output coupling loss for the 243 nm light with the BBO crystal, and an 18\% loss with the CLBO crystal due to the Fresnel reflection on the output facet of the crystal. The BBO crystal is cut at $\theta=$55$^\circ$ and has a double refraction angle of $\rho = 82$ mrad. For CLBO, $\rho = 16$ mrad and the crystal is cut with $\theta=$77$^\circ$.  As seen in Fig. \ref{fig:dub}, a Brewster oriented dichroic mirror with high reflectivity at 243 nm and high transmission at 486 nm is used to output couple the 243 nm radiation. The observed 243 nm output power as a function of the 486 nm input power is shown in Fig. \ref{fig:BBO} when using the BBO crystal and in Fig. \ref{fig:CLBO} for the CLBO crystal.  As can be seen from the figures, a greater efficiency was obtained with the CLBO crystal which we attribute to the smaller walkoff.  This also produced a 243 nm beam with better mode quality and less astigmatism.  

Due to the high UV power generated, crystal degradation is a concern. Frequency doubling studies at a similar wavelength have demonstrated 5 W of 266 nm power without damaging the CLBO crystal \cite{Sakuma:04}.  This corresponded to $\sim$ 5 times greater UV intensity within the crystal compared to the results reported here. Additional studies have shown that if degradation in CLBO occurs, it appears to be reversible \cite{Takachiho:14}. This is in contrast to BBO, which shows irreversible damage caused by the formation of an absorption center \cite{Scheid:07, Kondo:98, Takachiho:14}. Therefore, by utilizing CLBO we should be able to power scale our UV output as more fundamental power becomes available without crystal degradation.

CLBO is also known to be hygroscopic and some performance change has been reported as the crystal absorbs or desorbs water \cite{Kawamura:08}.  For this reason, the CLBO crystal was operated at a temperature of 130$^\circ$ C.  Over a few days at this elevated temperature, the conversion efficiency increased slightly above that shown in Fig. \ref{fig:CLBO} and we were able to observe $>$530 mW of 243 nm radiation over 50 minutes with no degradation.

\section{Conclusion}
\label{conclusion}
We have demonstrated a fiber based amplifier laser system capable of generating 6.3 W of power at 972 nm. Upon frequency doubling in successive resonant cavities, this laser source can generate 2.4 W at 486 nm and 530 mW at 243 nm. We are encouraged by the power scalability of our system.  It appears that our fiber amplifier platform should be able to produce additional 972 nm radiation through either more powerful, commercially available pump diodes or with an additional fiber amplifier stage of similar design \cite{Clarkson:10}.  To use the 915 nm pump radiation more efficiently would require that we obtain fibers with a larger core/cladding area ratio, such as the rod-type fibers used in \cite{Roser:08, Boullet:08}.  The doubling stages were designed with relatively loose focusing in the crystals.  This in conjunction with the high damage thresholds of LBO and CLBO make us hopeful that these cavities can also be power scaled.

Although we made no in-depth studies of the linewidth of our laser source for the work described here, our seed laser copies many aspects of the low phase noise design described in \cite{Kolachevsky:11} and we were able to couple our radiation into doubling cavities with few MHz resonance widths without any difficulty. For two-photon laser cooling of hydrogen, the transition width will be broadened to $\sim 50$ MHz by coupling the 2S and 2P states.  Therefore, the laser source we describe here already has the coherence necessary for that application.  That being said, spectroscopy of the hydrogen and anti-hydrogen 1S-2S transition with a power scaled 243 nm system would be very beneficial but would also require the source posses an extremely narrow linewidth.  Stabilizing this source to that level will therefore be the subject of future work.

\begin{acknowledgements}
We gratefully acknowledge Jacob Roberts for useful discussions and for carefully reviewing this manuscript.
\end{acknowledgements}
\bibliographystyle{spphys}       

\bibliography{243nmArXiv}
\end{document}